\documentclass[preprint,12pt]{elsarticle}

\usepackage{amsmath,amsfonts}
\usepackage{algorithmic}
\usepackage{graphicx}
\usepackage{float}
\usepackage{textcomp}
\usepackage{xcolor}
\usepackage{xspace}
\usepackage{fontawesome}
\usepackage{multirow}
\usepackage{rotating}
\usepackage{tikz}
\usepackage{soul}
\usepackage{makecell}
\usepackage{url}
\usepackage{enumitem}
\usepackage{booktabs}
\usepackage{fontawesome}
\usepackage{graphicx}
\usepackage{hyperref}
\usepackage{pifont}
\usepackage{multicol}
\usepackage{tcolorbox}
\usepackage{pifont}

\newcommand{\cmark}{\ding{51}}
\newcommand{\xmark}{\ding{55}}

\newcommand{\ie}{\emph{i.e.,}\xspace}
\newcommand{\eg}{\emph{e.g.,}\xspace}

\newcommand{\etal}{\emph{et~al.}\xspace}
\newcommand{\secref}[1]{Section~\ref{#1}\xspace}

\newcommand{\figref}[1]{Fig.~\ref{#1}\xspace}

\newcommand{\tabref}[1]{Table~\ref{#1}\xspace}

\newcommand{\approach}{LEONID\xspace} 
\newcommand{\java}{\emph{Java}\xspace}
\newcommand{\rev}[1]{\textcolor{black}{#1}}

\newcommand*\circled[1]{\tikz[baseline=(char.base)]{
		\node[shape=circle,fill,inner sep=0.8pt] (char) {\textcolor{white}{#1}};}}

\def\BibTeX{{\rm B\kern-.05em{\sc i\kern-.025em b}\kern-.08em
		T\kern-.1667em\lower.7ex\hbox{E}\kern-.125emX}}

\definecolor{gray50}{gray}{.5}
\definecolor{gray40}{gray}{.6}
\definecolor{gray30}{gray}{.7}
\definecolor{gray20}{gray}{.8}
\definecolor{gray10}{gray}{.9}
\definecolor{gray05}{gray}{.95}

\newlength\Linewidth
\def\findlength{\setlength\Linewidth\linewidth
	\addtolength\Linewidth{-4\fboxrule}
	\addtolength\Linewidth{-3\fboxsep}
}

\definecolor{lightergray}{rgb}{0.9,0.9,0.9}
\newtcolorbox{resultbox}{colback=lightergray, arc=0.5mm, top=2mm, bottom=2mm, left=2mm, right=2mm}

\begin{document}
\begin{frontmatter}

\title{Log Statements Generation via Deep Learning: Widening the Support Provided to Developers}

\author[usi]{Antonio Mastropaolo}
\ead{antonio.mastropaolo@usi.ch}

\author[usi]{Valentina Ferrari}
\ead{valentina.ferrari@usi.ch}

\author[eth]{Luca Pascarella}
\ead{lpascarella@ethz.ch}

\author[usi]{Gabriele Bavota}
\ead{gabriele.bavota@usi.ch}

\address[usi]{SEART @  Software Institute, Universit\`a della Svizzera italiana}
\address[eth]{Center for Project-Based Learning, ETH Zurich, Switzerland}

\begin{abstract}
	Logging assists in monitoring events that transpire during the execution of software. Previous research has highlighted the challenges confronted by developers when it comes to logging, including dilemmas such as where to log, what data to record, and which log level to employ (\eg info, fatal). In this context, we introduced LANCE, an approach rooted in deep learning (DL) that has demonstrated the ability to correctly inject a log statement into \java methods in $\sim$15\% of cases. Nevertheless, LANCE grapples with two primary constraints: (i) it presumes that a method necessitates the inclusion of logging statements and; (ii) it allows the injection of only a single (new) log statement, even in situations where the injection of multiple log statements might be essential. To address these limitations, we present \approach, a DL-based technique that can distinguish between methods that do and do not require the inclusion of log statements. Furthermore, \approach supports the injection of multiple log statements within a given method when necessary, and it also enhances LANCE's proficiency in generating meaningful log messages through the combination of DL and Information Retrieval (IR).
\end{abstract}

%

\begin{keyword}
Logging \sep DL for Software Engineering
\end{keyword}

\end{frontmatter}

\section{Introduction} \label{sec:intro}

The practice of injecting log statements in applications' code is widely adopted both in industry and open source projects \cite{oliner2012advances}. Indeed, log statements are instrumental to support several software-related activities, including program comprehension and debugging \cite{lu2017log,gurumdimma2016crude}. Given its popularity, it comes without surprise the proliferation of libraries to support logging activities: just for \java some possible options are Log4j \cite{log4j}, JCL \cite{jcl}, slf4j \cite{slf4j}, and logback \cite{logback}.

While logging is usually perceived as a good practice, it comes with its own drawbacks: Excessive logging could negatively impact performance and, if not carefully conceived, log statements can result in security issues such as providing access to user credentials or sensitive information. Also, researchers documented several bad practices that should be avoided while logging code \cite{Chen:icse2017,Li:icse2019}. 

In general, logging poses several challenges to software developers. First, they need to decide \emph{what to log}, by finding the right amount of log statements needed in the application without, however, flood it with useless log statements. Second, developers must \emph{log at the proper level}, namely select the proper log level for each entry (\eg info, warning, error). Third, log statements must be accompanied by \emph{meaningful and informative} log messages that can be easily understood. 

To support developers in these activities, researchers proposed techniques and tools automating specific aspects of logging, such as recommending (i) where/what to log \cite{yuan2010sherlog,jia2018smartlog,li2018studying,li2020shall}, and (ii) the right level to use for a given log statement \cite{yuan2012characterizing,oliner2012advances,li2017log,li2020qualitative,li2021deeplv}. In our first attempt to automate logging activities \cite{mastropaolo2022using}, we presented LANCE, an approach built on top of a Text-To-Text-Transfer-Transformer (T5) \cite{raffel2019exploring} deep learning (DL) model trained to generate and inject a complete log statement in a \java method provided as input. T5 has been pre-trained on a set of $\sim$6.8M \java methods using the classic ``masked language modeling'' objective \cite{raffel2019exploring}. In the case of LANCE, this means that during pre-training the model is provided as input a \java method with 15\% of its tokens masked and it is expected to predict the masked tokens. Such a pre-training task provides T5 with knowledge about the language of interest (\ie \java). 

Once pre-trained, the model has been fine-tuned for the specific task of interest. In this case, we selected $\sim$62k \java methods and removed from them exactly one log statement asking the model to generate and inject it, thus deciding \emph{where} to log (\ie in which part of the method), which \emph{log level} to use, and \emph{what} to log (\ie generate a meaningful log message in natural language). \rev{LANCE is the first approach supporting developers in all these activities. The empirical evaluation we run \cite{mastropaolo2022using} showed that LANCE was able to correctly predict the appropriate location of a log statement and its level in $\sim$66\% of cases, while the approach was struggling in predicting a meaningful log message, being successful in 15.2\% of test instances.}

While LANCE represents a step ahead in logging automation, it comes with some limitations. First, it assumes that only one log statement is needed in a \java method provided as input. This is due to the training procedure we employed that asks the model to always generate a single log statement. Second, given a \java method, LANCE cannot assess whether log statements are needed at all. Indeed, in some cases, enough log statements may be already present in the method or, maybe, the method does not feature statements that would benefit from logging. \rev{Finally, LANCE showed substantial limitations in synthesizing meaningful natural language log messages}. In this work we study how to partially address these limitations.

We start replicating LANCE by training and testing it on a dataset 3.6 times larger than the one we used originally  \cite{mastropaolo2022using} (230k training instances \emph{vs} 63k). Besides being larger, the new dataset features a more variegate set of log statements. Then, we present \approach as an extension of LANCE able to (i) discriminate between methods \emph{needing} and \emph{not needing} the injection of new log statements; and (ii) in case a need for log statements is identified, \approach, differently from LANCE, can decide the proper number of log statements to inject (which can be higher than one) and properly place them in the correct position.  \rev{We found that \approach can correctly predict the need for log statements with an accuracy higher than 90\%. Also, when log statements are needed, \approach can generate and inject in the right position multiple complete log statements in $\sim$17\% of cases.} 

Finally, in \approach we attempted to improve the performance achieved in the generation of meaningful log messages by exploiting a combination of DL and Information Retrieval (IR). Indeed, based on the results we achieved with LANCE, the generation of log messages really looked like the Achilles' heel of DL-based log generation. \rev{Results show that by increasing the size of the training dataset, the ability of LANCE in predicting meaningful log messages substantially improves (+100\% as compared to what we reported in \cite{mastropaolo2022using}). Instead, the combination of DL and IR we propose in \approach only marginally improves the results for this specific task (+5\% relative improvement).}

\begin{table*}[]
\centering
\scriptsize
 \caption{State-of-the-art approaches supporting developers in logging activities\vspace{-0.3cm}}
 \label{tab:sota}
 \resizebox{\textwidth}{!}{
				\begin{tabular}{lll|ccc|cc|c}
				
				\toprule
				\multirow{2}{*}{\textbf{Ref.}}                  & \multirow{2}{*}{\textbf{Venue}} & \multirow{2}{*}{\textbf{Name}} & \multicolumn{3}{c|}{\textbf{Log}} & \multicolumn{2}{c|}{\textbf{Log injection}} & \multirow{2}{*}{\textbf{Need for log statements}} \\
				                                                &                                 &                                & Level    & Position   & Message   & Single              & Multiple              &                                                  \\\midrule
				Zhu \etal~\cite{zhu2015learning}                & ICSE 2015                       & \textsc{LogAdvisor}            & \xmark   & \cmark     & \xmark    & \cmark              & \xmark                &  \xmark                                          \\
				Yao \etal~\cite{yao2018log4perf}                & ICPE 2018                       & \textsc{Log4Perf}              & \xmark   & \cmark     & \xmark    & \cmark              & \xmark                &  \cmark                                          \\
				Mizouchi \etal~\cite{mizouchi2019padla}         & ICPC 2019                       & \textsc{PADLA}                 & \cmark   & \xmark     & \xmark    & \cmark              & \cmark                &  \xmark                                          \\
				Li \etal~\cite{li2020shall}                     & ASE 2020                        & \textsc{}                      & \xmark   & \cmark     & \xmark    & \cmark              & \xmark                &  \xmark                                          \\
				Li \etal \cite{li2021deeplv}                    & ICSE 2021                       & \textsc{DeepLV}                & \cmark   & \cmark     & \xmark    & \cmark              & \xmark                &  \xmark                                          \\
				Ding \etal~\cite{ding2022logentext} 			& SANER 2022					  & \textit{LoGenText} 			   & \xmark   & \xmark     & \cmark    & \cmark				 & \xmark				 & 	\xmark											\\
				Mastropaolo \etal~\cite{mastropaolo2022using}   & ICSE 2022                       & \textsc{LANCE}                 & \cmark   & \cmark     & \cmark    & \cmark              & \xmark                &  \xmark                                          \\\midrule
				{\bf Our work}              & -                    & {\bf \textsc{LEONID}}                & \cmark   & \cmark     & \cmark    & \cmark              & \cmark                &  \cmark                                          \\
				\bottomrule
				\end{tabular}
}

\end{table*}

\tabref{tab:sota} shows how \approach widens the support provided to developers in the automation of logging activities. Indeed, it is the only one deciding whether log statements are needed in a method and, in case of positive answer, synthesizing multiple and complete log statements, and inject them in the correct position. 

\section{\approach} \label{sec:t5}

We start by providing an introduction to the T5 model we use (\secref{sub:t5}), the same we also exploited in LANCE \cite{mastropaolo2022using}. Then, we describe how we built the datasets used for the different training phases we deal with (\secref{sub:datasets}).  \secref{sec:training} will then explain how we used these datasets to run the actual training process.

\subsection{Text-to-Text-Transfer-Transformer (T5)}
\label{sub:t5}
T5 has been introduced by Raffel \etal \cite{raffel2019exploring} as a Transformer \cite{vaswani2017attention} model to support multitask learning. The idea behind T5 is to reframe NLP tasks in a unified text-to-text format in which the input and output of the model are text strings. The training of T5 includes two phases. The first is the \textit{pre-training}, in which the model is trained with a self-supervised objective to acquire general knowledge about the language(s) of interest. For example, this may mean providing as input to the model English sentences having a subset of their words masked and asking the model to generate as output the masked words. Being self-supervised (\ie the training instances can be automatically generated by masking random words) the pre-training can usually be performed on large-scale datasets. Once pre-trained, T5 can be fine-tuned to support specific tasks with supervised training objectives. This means, for example, providing it with pairs of sentences $<$\emph{english}, \emph{spanish}$>$ to train a translator.

In our work, we rely on the same T5 architecture (\ie T5$_{small}$) we exploited in LANCE \cite{mastropaolo2022using}. T5\textsubscript{\textit{small}} is characterized by six blocks for encoders and decoders. The feed-forward networks in each block consist of a dense layer with an output dimensionality ($d_{ff}$) of 2,048. The \textit{key} and \textit{value} matrices of all attention mechanisms have an inner dimensionality ($d_{kv}$) of 64, and all attention mechanisms have eight heads. All the other sub-layers and embeddings have a dimensionality ($d_{model}$) of 512. 
\rev{We acknowledge that employing larger models such as T5$_{base}$ or T5$_{large}$ can influence the performance of \approach when automating logging activities, but this comes at the expense of increased time and computational power requirements during the training process.}
The code implementing T5 is available in our replication package \cite{replication}.

\subsection{Datasets Needed for Training, Validation, and Testing}
\label{sub:datasets}

We start by describing the dataset used for pre-training T5 (\secref{sub:pretraining}). Then, we detail the several fine-tuning datasets we built (featuring training, validation, and test set). The first, aimed at replicating LANCE \cite{mastropaolo2022using}, teaches T5 how to inject a single log statement in a \java method  (\secref{sec:single-log-dataset}). The second fine-tuning dataset also focuses on the problem of injecting a single log statement, but this time exploits IR to provide T5 with concrete examples of log messages that might be relevant for the prediction at hand (\secref{sec:single-log-plus-IR}). This allows to compare LANCE with \approach in the task of single log statement injection. The third fine-tuning dataset trains \approach for the task of multi-log statements prediction, \ie injecting from 1 to $n$ log statements in a given method (\secref{sec:multi-log-dataset}). Finally, we describe the fine-tuning dataset to train a T5 able to discriminate between methods \emph{needing} and \emph{not needing} log statements (\secref{sec:predicting-dataset}). The datasets are summarized in Tables \ref{tab:ds-summary-1} and \ref{tab:ds-summary-2} and available in \cite{replication}.

All datasets have been built starting from the same set of GitHub repositories that we selected using the GHS (GitHub Search) tool by Dabi\'c \etal \cite{dabic2021sampling}. GHS allows to query GitHub for projects meeting specific criteria. We used the same selection criteria exploited in our former work on LANCE \cite{mastropaolo2022using}, selecting all public non-forked \java projects having at least 500 commits, 10 contributors, and 10 stars. These selection criteria aim at excluding personal/toy projects and reduce the chance of collecting duplicated code (non-forked repositories). We cloned the latest snapshot of the 6,352 projects returned by GHS. We scanned all cloned repositories to assess whether they featured a \texttt{POM} (Project Object Model) or a \texttt{build.gradle} file. Both these files allow to declare external dependencies towards libraries, the former using Maven, the latter Gradle. Such a check was performed since, as a subsequent step, we verify whether projects had a dependency towards Apache Log4j \cite{log4j} (\ie a well-known \java logging library) or SLF4J (Simple Logging Facade for \java) \cite{slf4j} (\ie an abstraction for \java logging frameworks similar to Log4j). Indeed, to train a T5 for the task of injecting complete log statement(s) in \java methods, we need examples of methods featuring log statements. The usage of popular logging \java libraries was thus a prerequisite for the project's selection.

We found 3,865 projects having either a \texttt{POM} or a \texttt{build.gradle} file and 2,978 of them featured a dependency towards at least one logging library. The overall projects' selection is very similar to the one we performed in \cite{mastropaolo2022using}, with the main differences being the additional mining of projects: (i) using Gradle as build system (in \cite{mastropaolo2022using} only Maven was considered); and (ii) having a dependency towards SLF4J (in \cite{mastropaolo2022using} only Log4j was considered). These choices help in increasing the size and variety of both the training and the testing datasets, making the prediction more challenging. 

We used srcML \cite{srcml} to extracted all \java methods in the selected projects. Then, we identified the log statements within each method (if any) and removed all methods featuring log statements exploiting custom log levels (\ie log levels that do not belong to any of the two libraries we consider, but that have been defined within a specific project). The valid log levels we considered are: \texttt{FATAL}, \texttt{ERROR}, \texttt{WARN}, \texttt{DEBUG}, \texttt{INFO}, and \texttt{TRACE}. At this point we were left with two sets of methods: those not having any log statement and those having at least one log statement using one of the ``valid'' log levels.

We run \emph{javalang} \cite{javalang} on these methods to tokenize them and excluded all those having $\#tokens < 10$ or $\#tokens \geq 512$. The upper-bound filtering has been done in previous works \cite{mastropaolo2021empirical,tufano2021automating,ciniselli2021empirical,tufano-mutants,Tufano:tosem2019} to limit the computational expenses of training DL-based models. The lower-bound of 10 tokens aims at removing empty methods. We also removed all methods containing non-ASCII characters in an attempt to exclude at least some of the methods featuring log messages not written in English. Finally, to avoid any possible overlap between the training, evaluation, and test datasets we are going to create from the collected set of methods, we removed all exact duplicates, obtaining the final set of 12,916,063 \java methods, of which 244,588 contain at least one log statement. 

\begin{table*}[h]
	\centering
	\scriptsize
	\caption{Number of methods in the datasets used in our study}
		\label{tab:ds-summary-1}
	\begin{tabular}{ccccccccc}
		\toprule
		\multirow{2}{*}{\textit{\textbf{Dataset}}} & \multicolumn{2}{c}{\textbf{train}} & \textbf{} & \textbf{eval} & \textbf{} & \textbf{test}  \\ \cline{2-3} \cline{5-5} \cline{7-7} 
		& \textbf{w/ log} & \textbf{w/o log} & \textbf{} & \textbf{w/ log} & \textbf{} & \textbf{w/ log} \\ \midrule
		\textit{Pre-training}              & -               &      12,671,475  &           & -               &           &  -               \\
		\textit{Fine-tuning: Single Log Generation}               & 229,703         & -                &           & 28,763          &           & 28,698          \\
		\textit{Fine-tuning: Single Log Generation with IR}               & 229,703         & -                &           & 28,763          &           & 28,698          \\
		\textit{Fine-tuning: Multi-log Injection with IR}               & 192,773         & -                &           & 24,092         &           & 24,088          \\
		\bottomrule
	\end{tabular}
\end{table*}

\subsubsection{Pre-Training Dataset}
\label{sub:pretraining}
Since the goal of pre-training is to provide T5 with general knowledge about the language of interest (\ie \java), we used for pre-training all methods not featuring a log statement (the latter will be used for the fine-tuning datasets). We adopted a classic \emph{masked language model} task, which consists in randomly masking 15\% of the tokens composing a training instance (\ie a \java method) asking the model to predict them.

\begin{figure}[h!]
	\centering
	\includegraphics[scale=0.35]{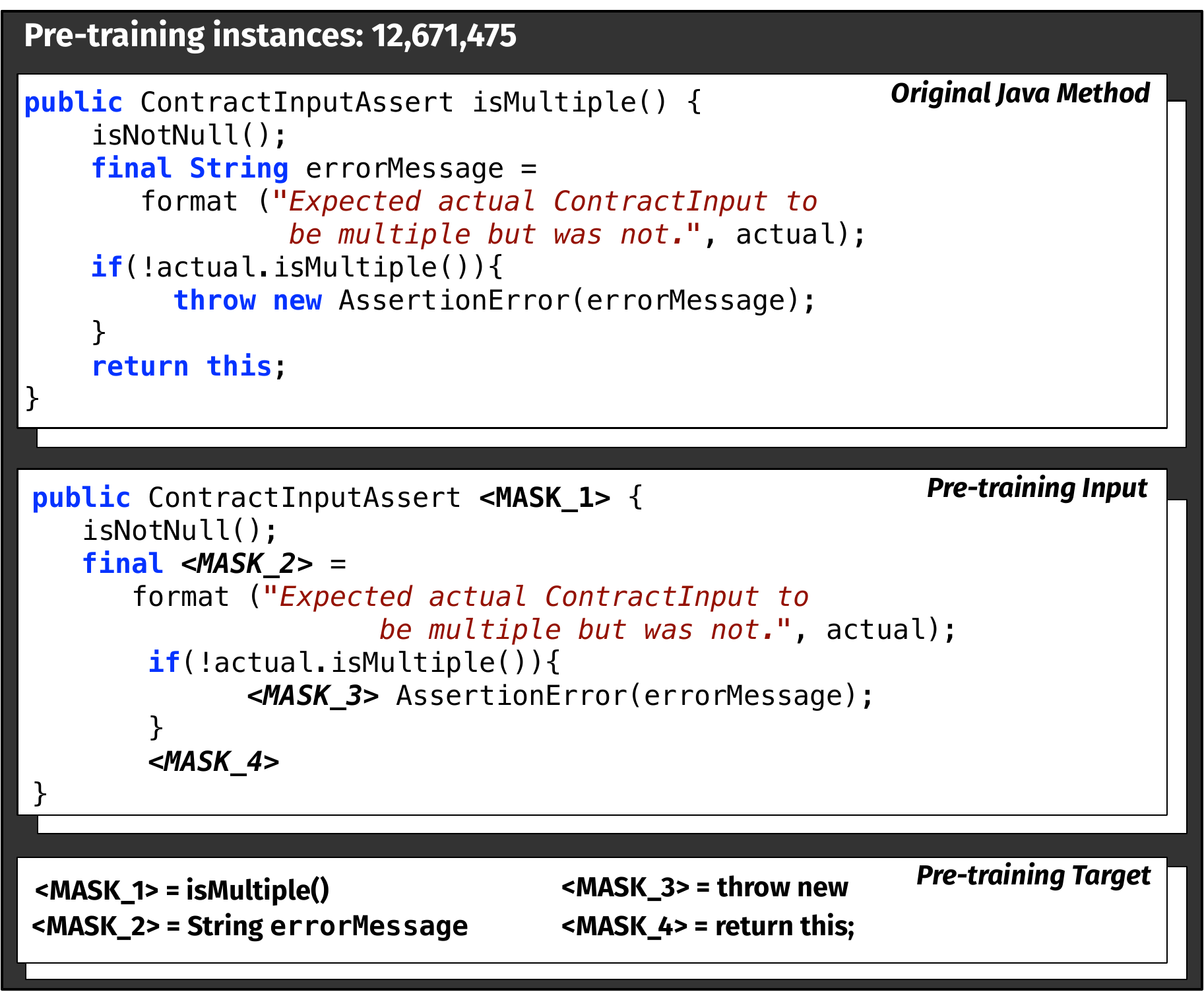}
	\caption{Example of Pre-training instance}
	\label{fig:pre-training}
\end{figure}

\figref{fig:pre-training} depicts the masking procedure of instances used to pre-train the model.

\begin{table*}[h!]
	\centering
	\scriptsize
	\caption{Number of methods in the datasets used to predict the need for log statements}
	\begin{tabular}{rcccccccc}
		\toprule
		\multirow{2}{*}{\textit{\textbf{Dataset}}} & \multicolumn{2}{c}{\textbf{train}} & \textbf{} & \multicolumn{2}{c}{\textbf{eval}}  & \textbf{} & \multicolumn{2}{c}{\textbf{test}}  \\ \cline{2-3} \cline{5-6} \cline{8-9} 
		& \textbf{Need} & \textbf{No need}   & \textbf{} & \textbf{Need} & \textbf{No need}   & \textbf{} & \textbf{Need} & \textbf{No need}   \\ \midrule
		\textit{Fine-tuning: Need4Log (50-50)}         & 98,848        & 92,126             &           & 12,257        & 11,468             &           & 11,627        &  11,627            \\
		\textit{Fine-tuning: Need4Log (75-25)}         & 98,848        & 92,126             &           & 12,257        & 11,468             &           & 12,159        &  4,053             \\
		\textit{Fine-tuning: Need4Log (25-75)}         & 98,848        & 92,126             &           & 12,257        & 11,468             &           & 3,875         &  11,627            \\ 
		\textit{Fine-tuning: Need4Log (2-98)}         & 98,848        & 92,126             &           & 12,257        & 11,468             &           & 238         &  11,627            \\ 
		\bottomrule
	\end{tabular}
	\label{tab:ds-summary-2}
\end{table*}

\subsubsection{Fine-tuning Dataset: Single Log Generation} \label{sec:single-log-dataset}
We build a fine-tuning dataset aimed at replicating what we did in the training of LANCE \cite{mastropaolo2022using}. We process each method $M$ having $n \geq 1$ log statements by removing from it one log statement (\ie leaving it with $n-1$ log statements). This allows to create a training pair $\langle M_s, M_t \rangle$ with $M_s$ representing the input provided to the model (\ie $M$ with one removed log statement) and  $M_t$ being the expected output (\ie $M$ in its original form, with all its log statements). This is the dataset used to train LANCE \cite{mastropaolo2022using} and it allows to train a model able, given a \java method as input, to inject in it one new log statement. For methods having $n > 1$ (\ie more than one log statement), we created $n$ pairs $\langle M_s, M_t \rangle$, each of them having one of the $n$ log statements removed (\ie different $M_s$). To ensure that after the log statement removal our instances still featured valid \java methods, we parsed each $M_s$ using JavaParser \cite{javaparser} and removed all pairs including an invalid $M_s$. 

We split the remaining pairs into training (80\%), validation (10\%) and test (10\%) set as reported in \tabref{tab:ds-summary-1}. Training and testing a T5 model on this dataset basically means performing a differentiated replication of LANCE on a 3.6$\times$ larger and more variegate (multiple logging libraries) dataset.

\subsubsection{Fine-tuning Dataset: Single Log Generation with IR} \label{sec:single-log-plus-IR}

In \approach, we combine DL and IR with the goal of boosting performance especially in the generation of meaningful log messages. The main idea is to augment the input provided to the model (\ie $M_{s}$) with log messages belonging to methods similar to $M_{s}$ which are featured in the training set. For each of the 244,588 $\langle M_s, M_t \rangle$ pairs in the fine-tuning dataset described in \secref{sec:single-log-dataset} (this includes training, validation, and test), we identify the $k$ most similar pairs in the training set. The similarity between two pairs is based on the similarity of their $M_s$ (\ie the method in which the log statement must be created) and it is computed using the Jaccard similarity \cite{hancock2004jaccard} index, based on the percentage of code tokens shared across the two methods. We then use these $k$ similar methods to extract from them examples of log messages used in coding contexts which are similar to the $M_s$ at hand. 

Two clarifications are needed. First, independently if a given pair is in the training, validation, or test set, we extract its $k$ most similar pairs only from the training set. This is needed since, while predicting the log statement to inject, the training set must be the only knowledge available to the model (\ie the test set must be composed of previously unseen instances). Second, when computing the Jaccard similarity, we remove from the compared methods all log statements, since we want to identify similar ``coding contexts'' that may require similar log statements. We created three different fine-tuning datasets using different values of $k=\{1,3,5\}$ (thus, a lower/higher number of exemplar log messages provided to the model).

\begin{figure}[h!]
	\centering
	\includegraphics[scale=0.35]{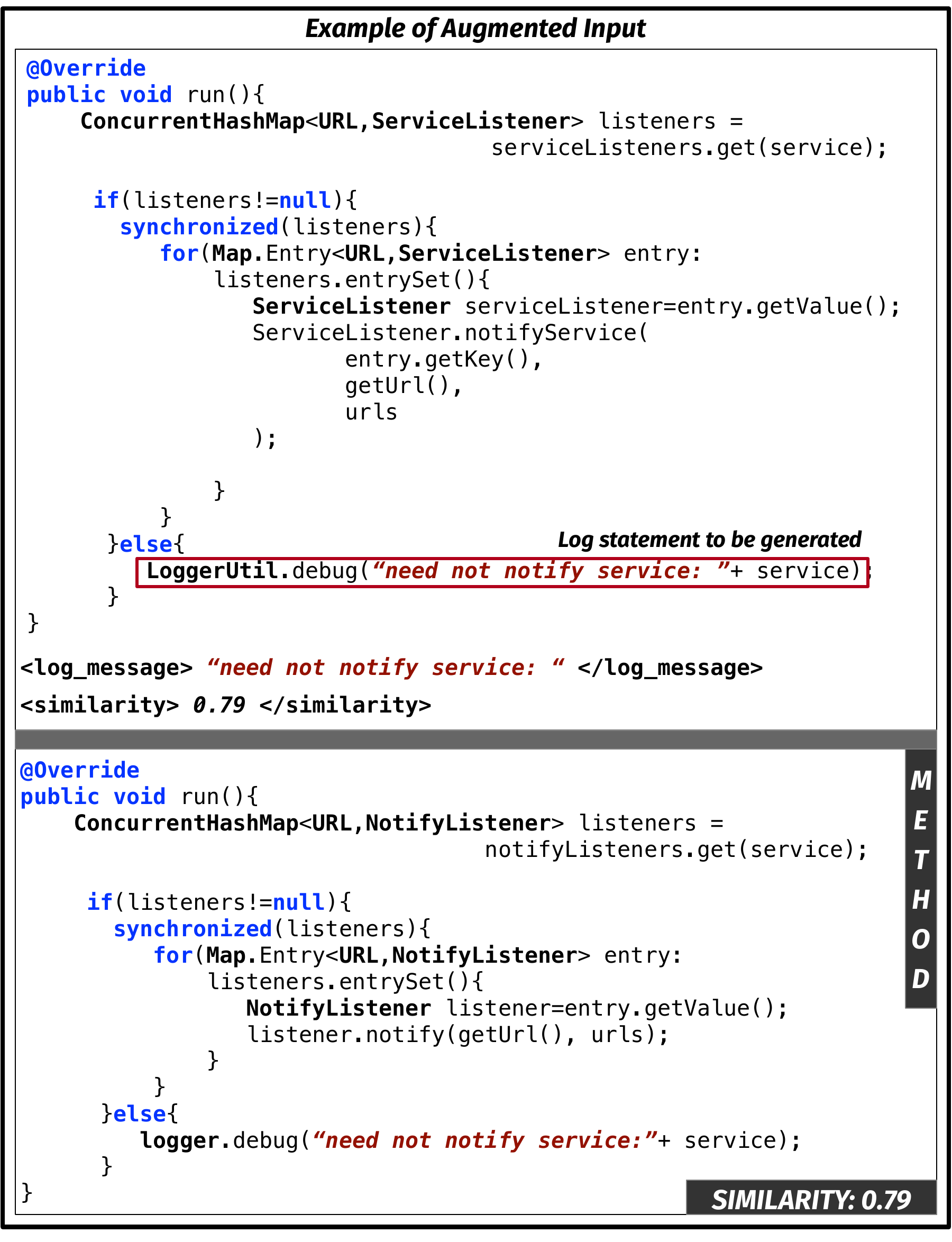}
	\caption{Example of instance in the ``Single Log Generation with IR'' dataset}
	\label{fig:ir-example}
\end{figure}

\figref{fig:ir-example} shows an example of training instance for this fine-tuning dataset. The method on top represents the $M_{s}$ \java method in which a log statement must be injected (\ie the one highlighted in red). The method is enriched with the exemplar log messages that have been found in the $k=1$ most similar method shown in the bottom. Besides the log messages, we also provide T5 with the Jaccard similarity between the $M_{s}$ at hand (top of the figure in this case) and the method of the training set from which the exemplar log message(s) has been extracted. This is meant to provide T5 with an additional hint in terms of which exemplar message comes from the most similar coding context (when more messages are retrieved).

Note that the instances in this dataset are exactly the same of the one previously described to replicate LANCE (see \tabref{tab:ds-summary-1}). This allows a direct comparison in terms of performance which will provide information about the gain, if any, provided by the IR integration.

\subsubsection{Fine-tuning Dataset: Multi-log Injection with IR} \label{sec:multi-log-dataset}

One limitation of LANCE \cite{mastropaolo2022using} we aim at addressing in this extension is the assumption that a \java method provided as input always requires one new log statement to be injected. 

Also for this dataset, \approach exploits a combination of DL and IR, thus we follow a process similar to the one described in \secref{sec:single-log-plus-IR}, with the main difference being the number of log statements we ask the model to generate. Given a method $M$ featuring $n$ log statements, we randomly select $y$ log statements to remove from it, with $1 \leq y \leq n$. This means that we create pairs $\langle M_s, M_t \rangle$ in which $M_s$ lacks a ``random'' number of log statements that must be generated by the model to obtain the target method $M_t$. This makes the prediction task substantially more challenging as compared to the single-log injection scenario experimented in LANCE. Also in this case we parsed each $M_s$ using JavaParser \cite{javaparser} and removed all pairs including an invalid $M_s$. The remaining part of the process (\ie identifying the $k$ most similar pairs to inject examples of log messages) is the same described in \secref{sec:single-log-plus-IR}. \tabref{tab:ds-summary-1} shows the distribution of instances among the training, evaluation, and test set for this dataset as well.

\subsubsection{Fine-tuning Dataset: Deciding Whether Log Statements are Needed} \label{sec:predicting-dataset}

While the dataset described in \secref{sec:multi-log-dataset} allows to build a model able to inject multiple log statements in a given \java method, such a model still assumes that at least one log statement must be injected in the input method. Thus, \approach also includes a T5 model trained as a binary classifier in charge of deciding whether a method provided as input requires the addition of log statements or not. In case of affirmative answer, the method can then be passed to the previously trained model which will decide how many and which log statements to inject.
To train such a classifier we again start from the original set of 244,588 \java methods having at least one log statement. Then, similarly to what done in \secref{sec:multi-log-dataset}, given a method $M$ featuring $n$ log statements, we randomly select $y$ log statements to remove from it with, however, $0 \leq y \leq n$. Thus, differently from the training dataset used for multi-log injection, we have instances from which we did not remove any log statement ($y=0$). Then, we create a pair $\langle M_s, B \rangle$ in which $M_s$ is the original method $M$ \textbf{possibly} lacking a random number of log statements, while $B$ is a boolean variable that could be equal \emph{true} (\ie $M_s$ needs the addition of log statements, since $y \geq 1$) or \emph{false} (\ie no log statements are needed in $M_s$, since $y = 0$). Non-parsable methods resulting after the removal of the log statements have then been removed, as well as duplicates resulting from different methods that, after the removal of log statements, become equal (\ie their only differences were the removed log statements). This process resulted in a dataset featuring 190,974 training instances (98,848 needing at least a log statement and 92,126 not needing it), accompanied by the evaluation and test sets summarized in \tabref{tab:ds-summary-2}.

As it can be seen, four different versions of the test set have been created, to experiment \approach in different scenarios. Let us explain such a choice. The test set should be representative of the real distribution of methods \emph{needing} and \emph{not needing} log statements. However, such a distribution cannot be computed in a reliable way. Indeed, one possibility we considered to build our dataset was to just consider all methods with and without log statements as training instances (as opposed to work only with methods having at least a log statement as we do). In a nutshell, the process would have been: (i) remove a random number of log statements from the methods with at least one log statement to create instances \emph{needing} logs; and (ii) assume that all methods without log statements do not require logging. However, assuming that all methods in a project not having log statements do not require logging is a very strong assumption. It is indeed possible that the project's developers just did not consider yet the usage of logs in a specific method or that, in a given project, logging is not yet a practice at all (thus all methods do not use log statements). This makes difficult a reliable computation of the number of methods \emph{needing} and \emph{not needing} logging. Also, such a problem justifies our decision to create instances of methods \emph{needing}/\emph{not needing} a log statement starting from all methods having at least one log statement and using the process described above (\ie removing a random number of statements to create instances in need of logging, and not removing any log statement to create instances not needing logging). At least, we are sure that these are methods for which developers considered logging (since they have at least one log statement) and, thus, can be seen as a sort of ``oracle''. 

The four test sets in \tabref{tab:ds-summary-2} simulate four different distributions of methods \emph{needing}/\emph{not needing} log statements: balanced (50\% per category), unbalanced towards \emph{needing} (75\%-25\%), unbalanced towards \emph{not needing} (25\%-75\%), and strongly unbalanced towards \emph{not needing} (2\%-98\%). 
The latter is a distribution we computed based on all 12M+ methods we mined, in which 98\% of methods do not have log statements, while 2\% have it. As said, this distribution is not completely reliable but, at least, gives an idea of what we found in the mined projects.

\section{Training and Hyperparameter Tuning} \label{sec:training}
All training we performed have been run using a Google Colab's 2x2, 8 cores TPU topology with a batch size of 128.

\subsection{Tokenizer Training}
Since we use software-specific corpora for pre-training and fine-tuning, we trained a tokenizer (\ie a SentencePiece model \cite{kudo2018sentencepiece}) on 1M \java methods randomly extracted from the pre-training dataset and 712,634 English sentences from the C4 dataset \cite{raffel2019exploring}. We included English sentences since, once fine-tuned, the models may be required to synthesize complex (natural language) log messages. We set the size of the vocabulary to 32k word-pieces.

\subsection{Pre-training}
We pre-trained T5 for 500k steps on the pre-training dataset composed by 12,671,475 \java methods (\tabref{tab:ds-summary-1}). Given the size of our dataset and the batch size, 500k steps correspond to $\sim$5 epochs. The maximum size of the input/output was set to 512 tokens.

\begin{table*}[h!]
	
	\centering
	\caption{T5 hyperparameter tuning results (in bold the best learning rate)}
\resizebox{\textwidth}{!}{
	\begin{tabular}{lrrrr}
		\toprule
		\textbf{Experiment}                  																		& \textbf{C-LR}              & \textbf{ST-LR}      & \textbf{ISQ-LR}        & \textbf{PD-LR} \\
		\midrule
		\textit{Fine-tuning: Single Log Generation with IR} ($k=1$)                         &   24.63\%                & 25.92\%    		           & \textbf{26.55\%}           &  26.36\%         \\
		\textit{Fine-tuning: Single Log Generation with IR} ($k=3$)                        &   26.25\%                & 26.04\%    		           & \textbf{26.68\%}          &  26.33\%         \\
		\textit{Fine-tuning: Single Log Generation with IR} ($k=5$)                         &  26.24\%                & 25.69\%    		           & \textbf{26.78\%}           &  26.33\%         \\
		\midrule
		\textit{Fine-tuning: Multi-log Generation with IR} ($k=1$)                         &   22.62\%                & 22.19\%    		           & \textbf{22.79\%}           &  22.76\%         \\
		\textit{Fine-tuning: Multi-log Generation with IR} ($k=3$)                        &   22.64\%                & 22.28\%    		           & \textbf{23.05\%}          &  22.59\%         \\
		\textit{Fine-tuning: Multi-log Generation with IR} ($k=5$)                         &   22.71\%                & 22.14\%    		           & \textbf{22.78\%}           &  22.51\%         \\
		\midrule
		\textit{Fine-tuning: Need4Log} & 96.58\%       & 96.56\%        & 96.59\%          & \textbf{96.62\%}\\
		\bottomrule
	\end{tabular}
}

	\label{tab:hp-results}

\end{table*}

\subsection{Hyperparameter Tuning}
Once pre-trained the model, we finetune the hyperparameters of the model following the same procedure we employed when developing LANCE. Such a procedure has been executed for each of the fine-tuning datasets previously described. In particular, we assessed the performance of T5 when using four different learning rate scheduler: (i) \textit{Constant Learning Rate} (C-LR): the learning rate is fixed during the whole training; (ii) \textit{Inverse Square Root Learning Rate} (ISR-LR): the learning rate decays as the inverse square root of the training step; (iii) \textit{Slanted Triangular Learning Rate \cite{howard2018universal}} (ST-LR): the learning rate first linearly increases and then linearly decays to the starting learning rate; and (iv) \textit{Polynomial Decay Learning Rate} (PD-LR): the learning rate decays polynomially from an initial value to an ending value in the given decay steps. 
The exact configuration of all the parameters used for each scheduling strategy is reported in \tabref{tab:learning-rates}.

\begin{table}[h]
	\centering
	\caption{Configurations for the experimented learning rates}
	\begin{tabular}{ll}
		\hline
		\textbf{Learning Rate Type} & \textbf{Parameters}               \\ \hline
		Constant                     & \textit{LR = 0.001}               \\
		Inverse Square Root         & \textit{LR\textsubscript{starting} = 0.01}  \\
		& \textit{Warmup = 10,000}          \\
		Slanted Triangular          & \textit{LR\textsubscript{starting} = 0.001} \\
		& \textit{LR\textsubscript{max} = 0.01}       \\
		& \textit{Ratio = 32}               \\
		& \textit{Cut = 0.1}                \\
		Polynomial Decay            & \textit{LR\textsubscript{starting} = 0.01}  \\
		& \textit{LR\textsubscript{end} = 0.001}      \\
		& \textit{Power = 0.5}              \\ \hline
	\end{tabular}
	\label{tab:learning-rates}
\end{table}

Each model has been run for 100k training steps on the fine-tuning dataset. Then, its performance has been assessed on the evaluation set in terms of correct predictions (\ie cases in which the generated output is equal to the target one). 

For the generative models injecting log statements this means that they outputted the \java method featuring all correct log statements in the expected positions. For the classifier, it means that it correctly predicted the need for log statements in a given method. The results achieved with each learning rate are reported in \tabref{tab:hp-results}. Our hyperparameter tuning required training and evaluating 28 models: For each of the 7 fine-tuning datasets in \tabref{tab:hp-results} we experimented 4 different learning rates. Given the achieved results, we will use the ISQ-LR for the generative models, and the PD-LR for the classifier when fine-tuning the models. Concerning the ``replication of LANCE'' (\ie fine-tuning T5 on the dataset \emph{Fine-tuning: Single Log Generation} in \tabref{tab:ds-summary-1}), we did not perform any hyperparameter tuning, but relied on the best configuration reported in the original paper \cite{mastropaolo2022using}, thus using the PD-LR.

\subsection{Fine-tuning}
Once identified the best learning rates to use, we fine-tuned the final models using early stopping, with checkpoints saved every 10k steps, a delta of 0.01, and a patience of 5. This means training the model on the fine-tuning dataset and evaluating its performance (again in terms of correct predictions) on the evaluation set every 10k. The training process stops if a gain lower than delta (0.01) is observed at each  50k steps interval. This means that after 60k steps, the performance of the model is compared against that of the 10k checkpoint and, if the gain in performance is lower than 0.01, the training stops and the best-performing checkpoint up to that training step is selected. This process has been used for all models, including the one replicating LANCE. Our replication package \cite{replication} reports the convergence of all models (\ie the steps after which the early stopping criterion was met).

\subsection{Generating Predictions}
Once the T5 models have been pre-trained and fine-tuned, they can be used to generate predictions for the targeted tasks. We generate predictions using a greedy decoding strategy, meaning that the generated prediction is the result of selecting at each decoding step the token with the highest probability of appearing in a specific position. Thus, a single prediction (\ie the one maximizing the likelihood of among all the produced tokens) is generated for an input sequence, as compared to strategies such as beam-search \cite{freitag2017beam} that generate multiple predictions.

\section{Study Design} \label{sec:design}

The \emph{goal} of our study is to evaluate the performance of \approach in supporting logging activities in \java methods. We focus on three scenarios: single log injection, in which we compare with our previous approach LANCE \cite{mastropaolo2022using}; multi-log injection; and deciding weather log statements are needed or not in a given \java method. The context is represented by the test datasets reported in \tabref{tab:ds-summary-1} (single and multi-log injection) and \tabref{tab:ds-summary-2} (deciding whether logging is needed).

We aim at answering the following research questions:

\begin{itemize}[itemindent=0.3cm]

\item[\textbf{RQ$_1$:}]\textit{To what extent is \approach able to correctly inject a single complete logging statement in \java methods?} RQ$_1$ mirrors the study we performed when presenting LANCE. We experiment \approach in the same scenario presented in \cite{mastropaolo2022using}: The injection of a single log statement in a given \java method. We compare the performance of \approach with that of LANCE when training and testing them on the same dataset. 

\item[\textbf{RQ$_2$:}]\textit{To what extent is \approach able to correctly inject multiple log statements when needed?} RQ$_2$ tests \approach in the more challenging scenario of injecting from 1 to $n$ log statements in a \java method, as needed.

\item[\textbf{RQ$_3$:}]\textit{To what extent is \approach able to properly decide when to inject log statements?} RQ$_3$ analyzes the accuracy of \approach in predicting whether or not log statements are needed in a given \java method. Additionally, we assess \approach as a whole using it to both predict the need for log statements and, subsequently, generate and inject them (if needed).
\end{itemize}

\subsection{Data Collection and Analysis}

To answer RQ$_1$ we run both \approach and LANCE against the test set described in \tabref{tab:ds-summary-1} for the single log generation task. The only difference is that LANCE has been trained on the dataset not featuring the exemplar log messages added through IR (row \emph{Fine-tuning: Single Log Generation} in \tabref{tab:ds-summary-1}), while \approach exploits this information (row \emph{Fine-tuning: Single Log Generation with IR} in \tabref{tab:ds-summary-1}). However, the training and test instances are exactly the same, allowing for a direct comparison. We assess the performance of the two techniques using the same evaluation schema employed in \cite{mastropaolo2022using}. In particular, we contrast the predictions generated by the two models against the expected output (\ie the \java method provided as input with the addition of the correct log statement). Note that generating and injecting a log statement  (\eg \texttt{LoggerUtil.debug("execution ok")}) involves correctly predicting several information: (i) the name of the variable used for the logging (\ie \texttt{LoggerUtil}); (ii) the log level (\ie \texttt{debug}); (iii) the log message (\ie \texttt{"execution ok"}); and (iv) the position in the method in which the log statement must be injected. Thus, when a prediction is generated, three scenarios are possible:

\textbf{Correct prediction:} A prediction that correctly captures all above-described information, \ie it matches the name used for the variable, the log level, message, and position as written by the original developers.

\textbf{Partially correct prediction:} A prediction that correctly captures a subset of the needed information (\eg it correctly generates the log statement but injects it in the wrong position).

\textbf{Wrong prediction:} None of the above-described information is correctly predicted.

We answer RQ$_1$ through the following combination of quantitative and qualitative analysis. On the quantitative side, we report for both \approach and LANCE the percentage of correct, partially correct, and wrong predictions. For the partially correct, we report the percentage of cases in which each of the ``log statement components'' (\ie variable name, log level, log message, and log position) has been correctly predicted. As for the percentage of correct and partially correct predictions, we pairwise compare them among the experimented techniques, using the McNemar's test \cite{mcnemar}, which is a proportion test suitable to pairwise compare dichotomous results of two different treatments. We complement the McNemar's test with the Odds Ratio (OR) effect size. We use the Holm's correction procedure \cite{Holm1979a} to account for multiple comparisons.

Concerning the quality of the log messages generated by the two techniques, looking for exact matches (\ie cases in which the generated log message is identical to the one written by developers) is quite limitative considering that a prediction including a message different but semantically equivalent to the target one could still be valuable. For this reason, we also compute the following four metrics used in Natural Language Processing (NLP) for the assessment of automatically generated text:

\textbf{BLEU}~\cite{papineni2002bleu} assesses the quality of the automatically generated text in terms of $n$-grams overlap with respect to the target text. The BLEU score ranges between 0 (the sequences are completely different) and 1 (the sequences are identical) and can be computed considering four different values of $n$ (\ie BLEU-\{1, 2, 3, 4\}). Besides these four variants, we also compute their geometric mean (\ie BLEU-A).

\textbf{METEOR}~\cite{meteor} is a metric based on the harmonic mean of unigram precision and recall. Compared to BLEU, METEOR uses stemming and synonyms matching to better reflect the human perception of sentences with similar meanings. Values range from 0 to 1, with 1 being a perfect match.

\textbf{ROUGE}~\cite{lin2004rouge} is a set of metrics focusing on automatic summarization tasks. We use the ROUGE-LCS (Longest Common Subsequence) variant which returns three values: the recall computed as \textit{LCS(X,Y)/length(X)}, the precision computed as \textit{LCS(X,Y)/length(Y)}, and the F-measure computed as the harmonic mean of recall and precision, where \textit{X} and \textit{Y} represent two sequences of tokens.

\textbf{LEVENSHTEIN Distance}~\cite{levenshtein1966} provides an indication of the percentage of words that must be changed in the synthesized log message to match the target log message. This is accomplished by computing the normalized token-level Levenshtein distance \cite{levenshtein1966} (NTLev) between the predicted log message and the target one. Such a metric can act as a proxy to estimate the effort required to a developer in fixing a non-perfect log message suggested by the model.

We also statistically compare the distribution of the BLEU-4 (computed at sentence level), METEOR, ROUGE, and LEVENSHTEIN distance related to the predictions generated by \approach and LANCE. We assume a significance level of 95\% and use the Wilcoxon signed-rank test \cite{wilcoxon}, adjusting $p$-values using the Holm's correction \cite{Holm1979a}. The  Cliff's Delta ($d$) is used as effect size \cite{Gris2005a} and it is considered: negligible for $|d| < 0.10$, small for $0.10 \le |d| < 0.33$, medium for $0.33 \le |d| < 0.474$, and large for $|d| \ge 0.474$ \cite{Gris2005a}.

On the qualitative side, we manually inspected 300 of the partially correct predictions generated by both techniques and having all information but the log message correctly predicted. The goal of the inspection is to verify whether the generated log message, while different from the target one, is semantically equivalent to it. To this aim, two of the authors independently inspected all 600 log messages (300 for each approach), with $\sim$11\% (70) arisen conflicts being solved by a third author. We report the percentage of ``wrong'' log messages generated by both techniques classified as semantically equivalent to the target one.

To answer RQ$_2$ and evaluate the extent to which \approach is able to correctly inject multiple log statements, we run \approach against the test set reported in \tabref{tab:ds-summary-1} (see row \emph{Fine-tuning: Multi-log Injection with IR}). We then report the percentage of correct predictions generated by the approach (\ie methods for which all $n$ log statements that \approach was supposed to generate and inject have been correctly predicted). In this case we do not compute the partially correct predictions since, if a prediction is not completely correct, it is not possible to match the generated log statements with the target ones to compare them. To make this concept more clear, consider the case in which \approach was asked to generate two log statements $s_1$ and $s_2$ but it only injects one statement $s_i$, being different from both $s_1$ and $s_2$. We cannot know whether $s_i$ should be compared with $s_1$ or with $s_2$ to assess the percentage of partially correct predictions in terms of \eg log level. For this reason, we only focus on the predictions being 100\% correct (\ie the output method is identical to the target one). 

To answer RQ$_3$, we run \approach against the test sets presented in \tabref{tab:ds-summary-2}, reporting the confusion matrix of the generated predictions and the corresponding accuracy, recall, and precision. We compare these results with those of: (i) an \emph{optimistic} classifier always predicting \emph{true} (\ie the method is in need for log statements); (ii) a \emph{pessimistic} classifier always predicting \emph{false} (\ie no need for log statements); and (iii) a random classifier, randomly predicting \emph{true} or \emph{false} for each input instance. We use the same statistical analysis described for RQ$_1$ to compare \approach with the baselines.
\section{Results Discussion} \label{sec:results}

We discuss the achieved results by research question.

\subsection{RQ$_{1}$: Injecting a single log statement}
\label{sec:rq1}

\tabref{tab:single-train-results} reports the results achieved by \approach and LANCE, in terms of correct and partially correct predictions for the task of single-log injection. For \approach we only report the results when $k=5$, since this is the variant that achieved the best performance (results with $k=1$ and $k=3$ are available in \cite{replication}). The first row of \tabref{tab:single-train-results} shows the percentage of correct predictions by both approaches, which is slightly higher for \approach (+1.8\% of relative improvement, from 26.78\% to 27.26\%). This difference is statistically significant (adj. $p$-value $<$ 0.01) with 1.12 higher odds of obtaining a correct prediction from \approach as compared to LANCE. 

\begin{table}[h!]
  \centering
  \scriptsize
  \caption{RQ$_1$: Correct and partially correct predictions by \approach and LANCE on the single-log injection task}
  \resizebox{\linewidth}{!}{
	  \begin{tabular}{cccccrrrr}
	  \toprule
	  Variable   & Level     & Message   & Position  &  & \multicolumn{1}{c}{LEONID (k=5)} & \multicolumn{1}{c}{LANCE} & \multicolumn{1}{c}{$p$-value} & \multicolumn{1}{c}{OR}     \\ \midrule
	  \ding{51}  & \ding{51} & \ding{51} & \ding{51} &  & 27.26\%        & 26.78\%                  &   $<$0.01                          & 1.12                            \\
	  \ding{51}  & -         & -         & -         &  & 76.45\%                    & 77.15\%                   &    $<$0.01             		 	& 0.88                            \\ 
	  -          & \ding{51} & -         & -         &  & 73.53\%                    & 74.18\%                   &   $<$0.01             			&  0.91                          \\
	  -          & -         & \ding{51} & -         &  & 31.55\%                    & 30.16\%                   &   $<$0.01                               &  1.36                         \\ 
	  -          & -         & -         & \ding{51} &  & 82.35\%                    & 82.28\%                  &   0.71                        &  1.01                           \\ \bottomrule
	
	  \end{tabular}
  }
  \label{tab:single-train-results}
\end{table}

The four subsequent rows report the cases in which one of the four log-statement components (variable, level, message, and position) was correctly predicted (\cmark), independently from whether the other three components were correct or not ($-$). As it can be seen, there is no significant difference in the prediction of the log position, with both techniques correctly predicting it in $\sim$82.3\% of cases. Differences are observed for the log variable and level in favor of LANCE (+1.0\% and +0.9\% relative improvement), and for the log message in favor of \approach (+4.6\% relative improvement). The log message is the part for which we observed the highest OR among all comparisons. Considering that the only difference between \approach and LANCE is the usage of IR, the improvement in the generation of meaningful log messages we targeted has been at least partially achieved. The latter has, however, a small price to pay in the correct prediction of the log variable and level. Still, for these elements \approach is able to generate a correct prediction in over 73.5\% of cases, while the correct generation of the log message still represents the Achilles' heel of these techniques, with 31.55\% correct predictions achieved by \approach. Thus, we believe that improvements on the log message predictions should be favored even at the expense of losing a bit of prediction capabilities on other elements. 

Digging further into the quality of the generated log messages, \tabref{tab:log-messages-stats} reports the results computed using the four NLP metrics presented in \secref{sec:design} for both models (in bold the best results). All metrics suggest that the log messages generated by \approach are closer to those written by humans. According to our statistical analysis (results in \tabref{tab:test-wilcoxon}), all these differences are statistically significant (adj. $p$-value $<$ 0.001) with, however, a negligible effect size. 

\begin{table}[h]
	\centering
	\caption{RQ$_1$: Evaluation Metrics on Log Messages: LEONID \emph{vs} LANCE}
	\scriptsize
	\label{tab:log-messages-stats}
	\begin{tabular}{lrr}
		\toprule
		& {\bf LANCE}  &  {\bf LEONID ($k=5$)} \\\midrule
		BLEU-A \cite{papineni2002bleu}& 31.98 & \bf 35.36\\
			\hspace{0.2cm} BLEU-1 & 47.30  & \bf 50.00\\
			\hspace{0.2cm} BLEU-2 & 36.30  & \bf 39.60\\
			\hspace{0.2cm} BLEU-3 & 33.90  & \bf 35.00\\
			\hspace{0.2cm} BLEU-4 & 31.40  & \bf 32.40\\
		METEOR \cite{meteor} & 58.60  & \bf 60.35 \\
		ROUGE-LCS \cite{lin2004rouge} &  \\
		\hspace{0.2cm} $precision$ & 42.57 & \bf 44.68\\
		\hspace{0.2cm} $recall$ & 44.04 &   \bf 46.01\\
		\hspace{0.2cm} $fmeasure$ & 42.19 &  \bf 44.33\\
		LEVENSHTEIN \cite{levenshtein1966} & 44.02 & \bf 41.85 \\\bottomrule
	\end{tabular} 
\end{table}

\begin{figure}[!ht]
	\centering
	\includegraphics[scale=0.35]{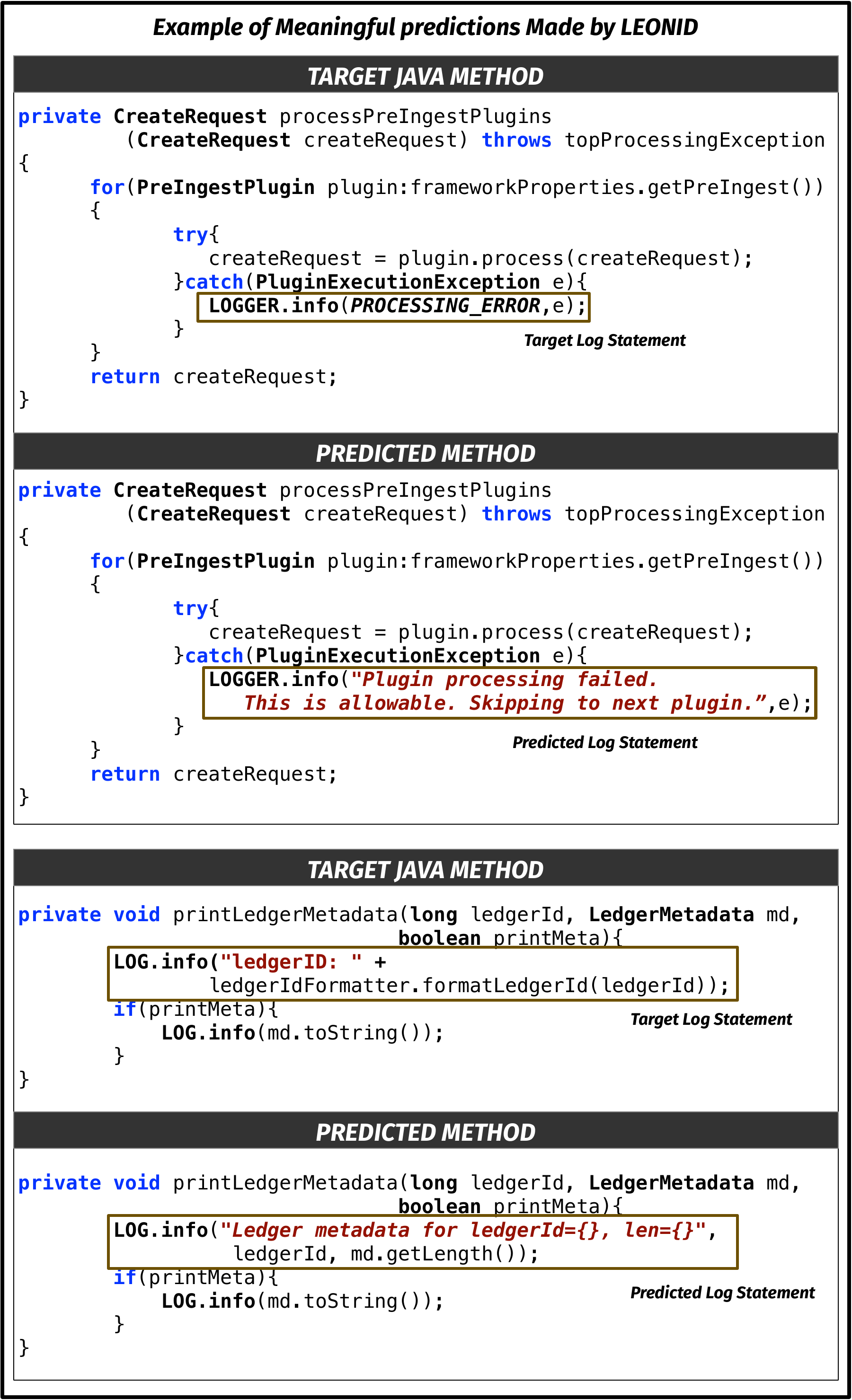}
	\caption{Examples of semantically equivalent log messages generated by \approach}
	\label{fig:equivalent-messages}
\end{figure}

\begin{table}[ht]
	\centering
	\caption{RQ$_1$: Statistical Tests: LEONID \emph{vs} LANCE for NLP metrics}
	\scriptsize
	\label{tab:test-wilcoxon}
	\begin{tabular}{llrc}
		\toprule
		\textbf{Comparison} & \textbf{Metric} & \textbf{\emph{p}-value} & \textbf{d} \\ 
		\midrule
		\multirow{4}{*}{LEONID ($k=1$) \emph{vs.} LANCE} & BLEU-4 & $<$0.001 & -0.022 (N) \\ 
		& METEOR & $<$0.001 & -0.025 (N) \\
		& ROUGE-LCS (f-measure) & $<$0.001 & -0.025 (N) \\ 
		& LEVENSHTEIN & $<$0.001 & +0.022 (N) \\\midrule
		\multirow{4}{*}{{LEONID ($k=3$) \emph{vs.} LANCE} } & BLEU-4 & $<$0.001 & -0.026 (N) \\ 
		& METEOR & $<$0.001 & -0.029 (N) \\
		& ROUGE-LCS (f-measure) & $<$0.001 & -0.023 (N) \\ 
		& LEVENSHTEIN & $<$0.001 & +0.027 (N) \\\midrule
		\multirow{4}{*}{{LEONID ($k=5$) \emph{vs.} LANCE} } & BLEU-4 & $<$0.001 & -0.026 (N) \\ 
		& METEOR & $<$0.001 & -0.029 (N) \\
		& ROUGE-LCS (f-measure) & $<$0.001 & -0.026 (N) \\ 
		& LEVENSHTEIN & $<$0.001 & +0.029 (N) \\\midrule
	\end{tabular}
\end{table}

Also the result of our manual inspection of 300 partially correct predictions by \approach and by LANCE point to a similar story: We found 198 of those generated by \approach (66\%) to report the same information of the target log message, despite being semantically different. The remaining 102 (34\%) predictions, instead, reported a log message completely different from the target one or not meaningful at all. For LANCE, the number of semantically equivalent log messages is slightly lower --- 192 (64\%) --- but inline with that observed for \approach. Examples of different but semantically equivalent log messages generated by \approach are reported in \figref{fig:equivalent-messages}. The methods labeled with ``Target Java Method'' represent the ``oracle'', namely the log statement that \approach was supposed to generate. Those instead labeled with ``Predicted Method'' represents the generated prediction being different from the expected target but, accordingly to our manual analysis, still valid.

\vspace{0.2cm}
\begin{resultbox}
\textbf{Answer to RQ$_1$.} The 3.6 larger training dataset (as compared to the original one we used in \cite{mastropaolo2022using}), resulted in a boost of performance when predicting the log message (15.20\% in \cite{mastropaolo2022using} \emph{vs} 30.16\%). Such a result has been further improved by \approach, which achieves a +4.6\% relative improvement (\ie 31.55\% of correctly generated log messages). All metrics used to assess the quality of the log messages generated by \approach indicate improvements over LANCE. However, these improvements are marginal, showing that more research is needed to further improve the automated generation of log messages.
\end{resultbox}

\subsection{RQ$_{2}$: Injecting multiple log statements}
\label{sec:rq2}
\begin{figure}[ht!]
	\centering
	\includegraphics[scale=0.35]{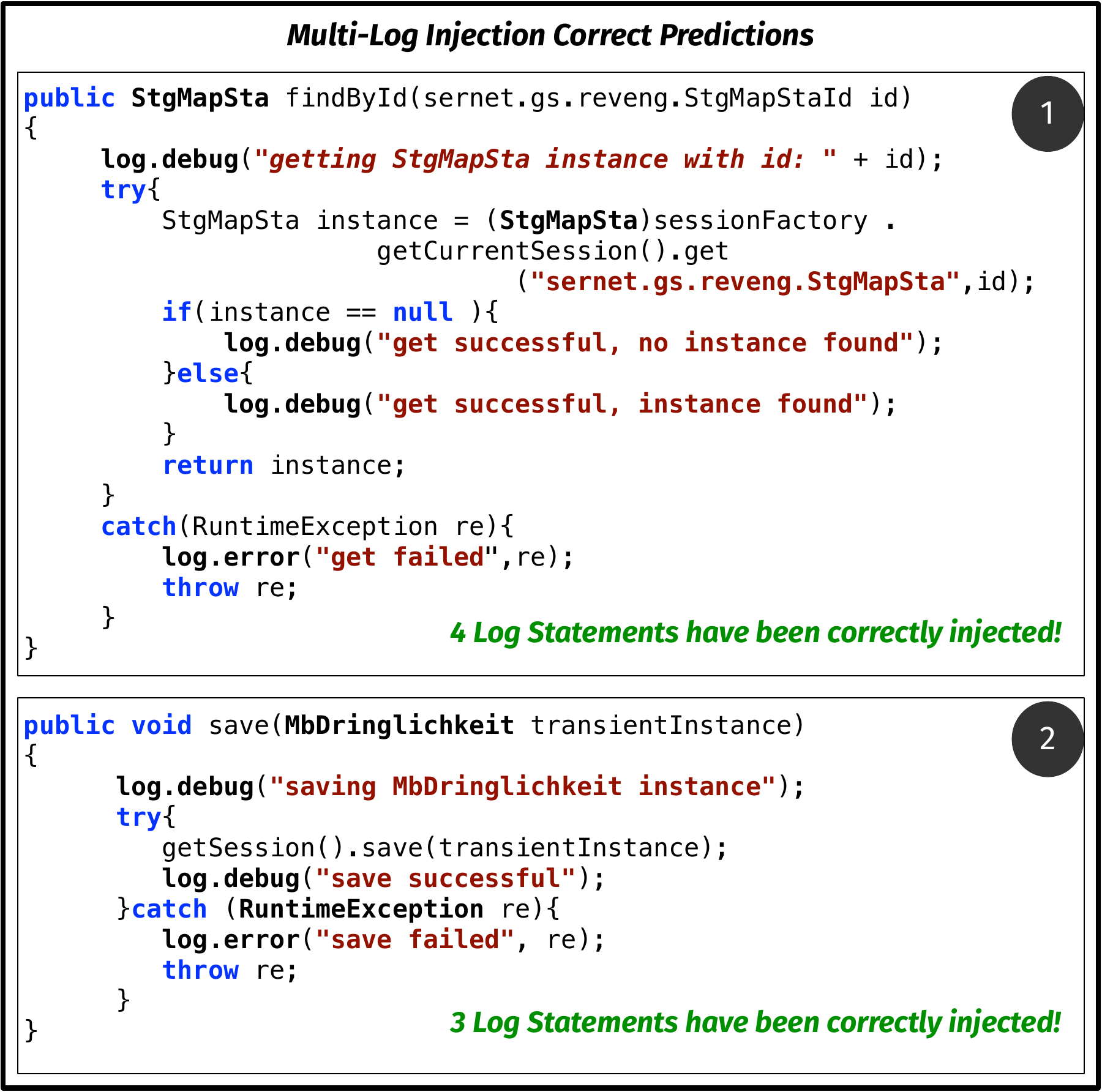}
	\caption{Correct predictions made by \approach when injecting more than one log statement.}
	\vspace{0.4cm}
	\label{fig:multi-leonid}
\end{figure}

As explained in \secref{sec:design}, it is not possible to compute the partially correct predictions in the scenario of multiple log injection. Thus, we limit our discussion to the correct predictions generated by \approach. Independently from the value of $k$ (\ie the number of similar coding contexts from which exemplar log messages are extracted), \approach can correctly predict all log statements to inject in a given method in $>$23\% of cases. Also in this scenario, $k=5$ is confirmed as the best configuration, with 23.51\% of correct predictions. \figref{fig:multi-leonid} depicts two cases for which \approach correctly recommended more than one log statement: \textit{four} in \circled{1} and \textit{three} in \circled{2}.

Interestingly, the drop in performance as compared to the simpler scenario of single log injection is there but is not substantial (27.26\% \emph{vs} 23.51\%). Remember that in this experiment we removed from a given \java method $M$ a random number $y$ of log statements, with $1 \leq y \leq n$ and $n$ being the number of log statements in $M$. Thus, it is possible that most of the methods in our dataset had $n=1$ and, as a consequence, $y=1$ (\ie \approach must generate one log statement), thus making the task similar to the single-log injection. For this reason, we inspected our test set and found indeed that 85\% of methods in it featured, in their original form, a single log statement. On top of this, there is another 6.7\% of methods which originally had more than one log statement and from which we randomly removed $y=1$ statement, thus again resulting in instances requiring the addition of a single log statement. We clustered the instances in the test set based on the number of log statements that \approach was required to generate. We created two subsets: (i) \emph{one-log}, having $y=1$; and (ii) \emph{at-least-two-log}, $y\geq2$. The \emph{one-log} subset features 91.7\% of the instances in the test set (22,104 out of 24,088) and, on those, \approach achieves 24.1\% correct predictions; the \emph{two-log} subset features 1,984 instances (8.3\%), on which \approach has a 17.0\% success rate. Thus, there is an actual performance drop when \approach needs to predict multiple log statements in a given method. Still, in 17\% of cases, \approach is able to inject the same log statements manually written by developers. To give a term of comparison, in our original paper presenting LANCE \cite{mastropaolo2022using}, we reported a 15.2\% success rate for the task of single-log injection.

\vspace{0.2cm}
\begin{resultbox}
\textbf{Answer to RQ$_2$.} \approach can support the task of multiple log injection, achieving 17.0\% of correct predictions when more than one log statement must be injected. It is important to highlight that in this task it is up to the model to infer how many log statements are actually needed in the method given as input, making it more complex than the single-log injection experiment even when only a single log statement must be injected.
\end{resultbox}

\subsection{RQ$_{3}$: Deciding whether log statements are needed}
\label{sec:rq3}
\begin{figure}[h!]
	\centering
	\includegraphics[scale=0.31]{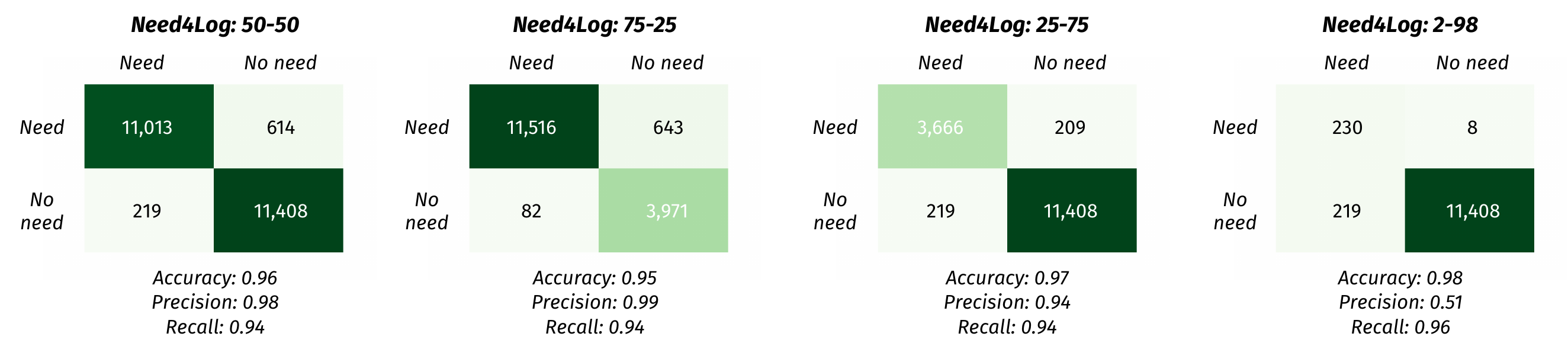}
	\caption{RQ$_3$: Results achieved by \approach when deciding whether log statements are needed or not in \java methods}
	\label{fig:rq3-cm}
\end{figure}

\figref{fig:rq3-cm} reports the confusion matrices for the test sets in \tabref{tab:ds-summary-2}, differing for the proportion of \emph{need}/\emph{no need} instances they feature. 
The rows in the matrices represent the oracle and columns the predictions. For example, the first matrix to the left indicates that out of the 11,627 (11,013+614) methods in \emph{need} for log statements, \approach correctly identified 11,013 of them, wrongly reporting the remaining 614 as \emph{no need}.

The overall accuracy of the classifier is always very high ($\geq$0.95), indicating that most of instances are correctly classified. Similarly, the recall for the ``need'' class is always $\geq$0.94 (see \figref{fig:rq3-cm}), suggesting that most of the methods in \emph{need} of log statements are identified. 

Instead, the precision drops to 0.51 when the test set is very unbalanced towards the ``no need'' class, with only 238 \emph{need} instances. Indeed, every classification error weights a lot more on the precision when the number of \emph{need} instances is so low: The 219 misclassifications represent 49\% --- 219/(230+219) --- of the instances that \approach classifies as in \emph{need} of log statements. Given the overall very good performance achieved by \approach, we decided to inspect these 219 instances to understand the rationale behind the recommendation by \approach (\ie add log statements). What we found is that, indeed, these are cases which are worth the attention of the developers since they may benefit from additional logging.

\begin{figure}[ht!]
	\centering
	\includegraphics[scale=0.35]{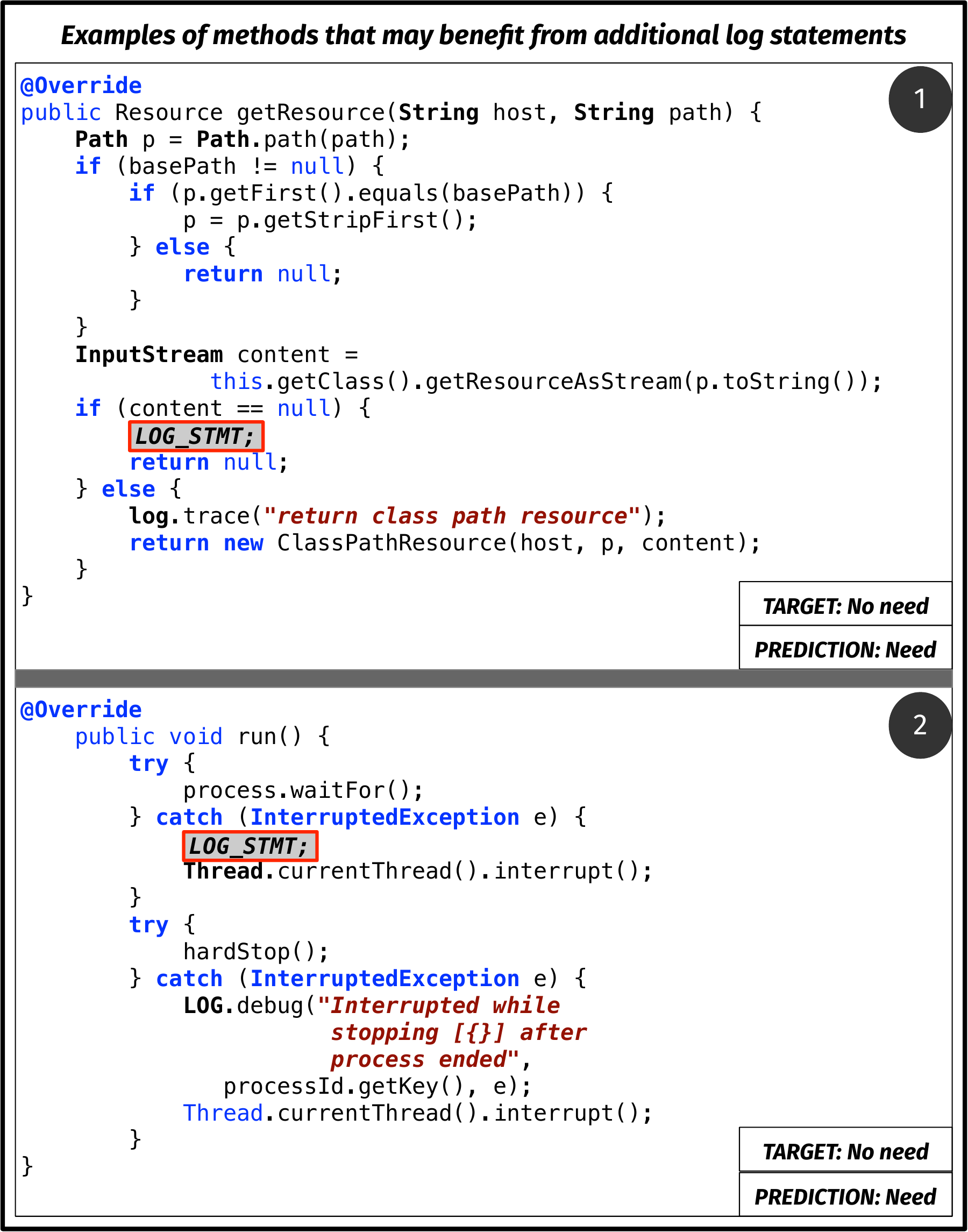}
	\caption{$RQ_{3}$: Examples of methods that may benefit from further logging}
	\label{fig:no-need}
\end{figure}

\figref{fig:no-need} shows two examples of ``\emph{no need} methods'' classified by \approach as in \emph{need} for additional log statements. We added the \emph{LOG\_STMT} text bordered in red to indicate positions which may benefit of logging, especially considering the other log statements present in the method. For example, in method \texttt{run} \circled{2} the developers used a log statement to document the reason for the \texttt{InterruptedException} in the second \texttt{try/catch}, while a similar scenario in the first \texttt{try/catch} is not logged. Overall, based on our manual inspection of the ``false positives'', we are confident that these could still represent valuable recommendations for developers.

When comparing the correct predictions achieved by \approach with those of the optimistic, pessimistic, and random classifier, we always found a statistically significant difference in favor of \approach (adj. $p$-value $<$ 0.001) accompanied by an OR going from a minimum of 6.17 to a maximum of 1,426. The only exception is, as expected, the comparison with the pessimistic classifier on the 2-98 test set, on which the pessimistic classifier achieves 98\% of correct predictions. In this case, we found no statistically significant difference (adj. $p$-value $=$ 0.63) with \approach (detailed results in \cite{replication}).

\begin{figure}[ht!]
	\centering
	\includegraphics[width=\columnwidth]{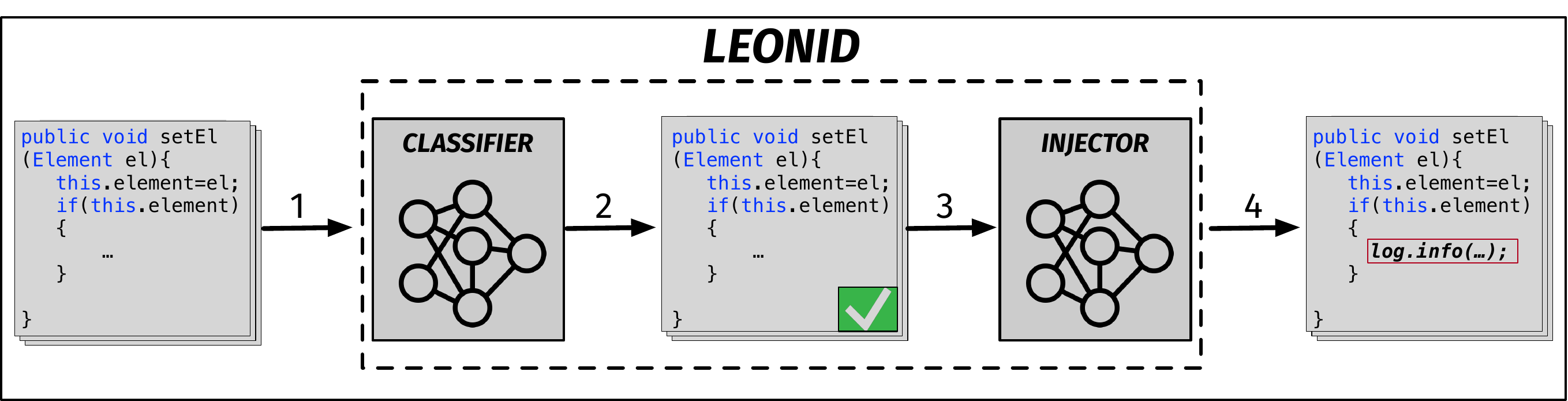}
	\caption{$RQ_{3}$: Example of \approach operating in an end-to-end logging scenario (\ie classification and injection).}
	\label{fig:approach}
\end{figure}

Finally, we conducted a full-system assessment in which we integrated the classifier and generator into a pipeline that first determines whether log statements are necessary, and if so, the module responsible for injecting the logs is activated.  \figref{fig:approach} provides an overview of how \approach operates in an \texttt{end-to-end} logging scenario. In this context, the \texttt{CLASSIFIER} module first determines whether log statements are required for the target method. If log statements are necessary, the \texttt{INJECTOR} component inserts one or more log statements into the provided \java method.

The achieved results showed that our \texttt{end-to-end} logging system can correctly inject $\sim$23\% (5,538/24,088) log statements when needed. This must be compared with the 27.26\% achieved in RQ$_1$ when we only assessed the generation of log statements, ``providing'' \approach only with instances that needed a log statement. Thus, while there is a slight loss in performance, the achieved results confirm the ability of \approach in automatically assessing the need for log statements. 

\begin{resultbox}
\textbf{Answer to RQ$_3$.} \approach can discriminate between methods \emph{needing} and \emph{not needing} additional log statements, with an accuracy higher than 0.95. This allows \approach to both predict the need for log statements and generating them.

\end{resultbox}

\section{Threats to Validity} \label{sec:threats}

\textbf{Construct validity.} The building of our fine-tuning datasets rely on the assumption that the exploited code instances, as written by developers, represent the ``correct'' predictions that the models should generate. This is especially true for the classifier aimed at predicting whether log statements are needed. For example, the instances that we labeled as ``\emph{not needing log statements}'' are methods featuring $n \geq 1$ log statements from which we did not remove any log statement. Thus, we assume that these methods need exactly $n$ log statements (\ie the ones injected by the developers), not one more. This is a strong assumption, as confirmed by the examples in \figref{fig:no-need}. 

\rev{In addition, there is evidence in the literature showing that some projects may adopt suboptimal logging practices \cite{patel2022sense}, thus again posing question on the quality of the adopted ground truth. Future work should involve developers in the assessment of the recommendations generated by \approach or similar techniques.
}
Still, using the code written by developers as oracle is a popular practice in DL for SE \cite{tufano2022using, Tufano:tosem2019, tufano-mutants, watson2020learning, tufano2022generating}.

\rev{It is important to notice that, when preparing the fine-tuning datase we removed log statements from any location within a \java method. As a consequence, certain methods may contain empty blocks (\eg an empty \texttt{if} block that only contained the log statemet), thus hinting the model to the right location in which the log statement should be injected (since there is likely something missing in that unusual empty block). To address this problem, we assessed the model's performance on a subset of our initial test set featuring 17,455 instances ($\sim$73\% of the original test set) in which there were no empty blocks left within the test method after removing the log statements. The results indicate that \approach remains competitive even in this more challenging scenario, correctly generating and injecting log statements in 25.30\% (4,416/17,455) of the test instances (as compared to the 27.26\% obtained on the full test set).}

\textbf{Internal validity.}  We performed a limited hyperparameters tuning only focused on identifying the best learning rate, while we relied on the best architecture identified by Raffel \etal \cite{raffel2019exploring} for the other parameters. We acknowledge that additional tuning can result in improved performance. \rev{
	Also, different similarity measures used to retrieve similar $M_s$ from the training set may lead to different results. Our choice of the Jaccard similarity was due to practical reasons: Since a given input method to \approach must be compared with all entries in the training set, we needed a very efficient similarity measure in terms of required computational time. For example, we also implemented a variant of \approach exploiting CodeBLEU \cite{ren2020codebleu} as a similarity measure. Considering that larger and larger training sets will be likely used in the future, a scalable solution is a must also to make \approach usable in practice.
}

\textbf{External validity.} Our research questions have been answered using a dataset being 3.6 times larger as compared to the dataset we originally used when proposing LANCE \cite{mastropaolo2022using}. Also, the new dataset is more variegated, featuring projects using different build systems (as compared to the Maven-only policy we relied in \cite{mastropaolo2022using}) and having dependencies towards different logging libraries (differently from the original Log4j-only policy we end up using in \cite{mastropaolo2022using}). Still, we do not claim generalizability of our findings for different populations of projects, especially those written in other programming languages. \rev{
	This holds not only when looking at the performance achieved on our test set (\ie different test sets can yield to different results), but also when considering the usage in \approach of information collected via IR from the training set (\ie the performance observed for \approach are bounded to the variety of data present in our training set). Additional experiments are needed to corroborate/contradict our findings.
}
\section{Related Work} \label{sec:related}

\subsection{Empirical Studies on Logging Practices}
Yuan \etal \cite{yuan2012characterizing} conducted one of the first empirical study on logging practices in open-source systems, analyzing C and C++ projects. They show that developers make massive usage of log statements and continuously evolve them with the goal of improving debugging and maintenance activities.

Fu \etal \cite{fu2014developers} studied the logging practices in two industrial projects at Microsoft, investigating in particular which code blocks are typically logged. They also propose a tool to predict the need for a new log statement, reporting a 90\% F-Score.

Chen \cite{chen2017characterizing} and Zeng \etal \cite{zeng2019studying} extended the study of Yuan \etal \cite{yuan2012characterizing} to \java and Android systems, respectively. In particular, Chen analyzed 21 Java-based open-source projects while Zeng \etal considered 1,444 open-source Android apps mined from F-Droid. Both studies confirmed the results of Yuan \etal \cite{yuan2012characterizing}, finding a massive presence of log statements in the analyzed systems. 

Zhi \etal \cite{zhi2019exploratory} investigated how logging configurations are used and evolve, distilling 10 findings about practices adopted in logging management, storage, formatting, and configuration quality. Other researchers studied the evolution and stability  of log statements. For example, Kabinna \etal \cite{kabinna2018examining} examined how developers of four open source applications evolve log statements. They found that nearly 20-45\% of log statements change throughout the software lifetime. 

Zhou \etal \cite{zhou2020mobilogleak} explored the impact of logging practices on data leakage in mobile apps. In addition, they propose MobiLogLeak to automatically identify log statements in deployed apps that leak sensitive data. Their study show that 4\% of the analyzed apps leak sensitive data.

Recently, Li \etal~\cite{li2020qualitative} conducted an extensive investigation on logging practice from a developer's perspective. The goal of this research is to push the design of automated tools based on actual developers' needs (rather than on researchers' intuition). The authors surveyed 66 developers and analyzed 223 logging-related issue reports shedding light on the trade-off between costs and benefits of logging practices in open source. The results show that developers adopt an \emph{ad hoc} strategy to compensate costs and benefits while inserting logging statements for various activities (\eg debugging). 

The above-described papers lay the empirical foundations for techniques supporting developers in logging activities (including our work). Approaches such as \approach can help in reducing the cost of logging while supporting developers in taking proper decisions when they wish to add log statements.

\subsection{Automating Logging Activities}

Researchers proposed techniques and tools to support developers in logging activities.

\textbf{Log message enhancement.} Yuan \etal \cite{yuan2012improving} proposed \textsc{LogEnhancer} as a prototype to automatically recommend relevant variable values for each log statement, refactoring its message to include such values. Their evaluation on eight systems demonstrates that \textsc{LogEnhancer} can dramatically reduce the set of potential root failure causes when inspecting log messages. Liu \etal \cite{liu2019variables} tackled the same problem using, however, a customized deep learning network. Their evaluation showed that the mean average precision of their approach is over 84\%.

Ding \etal proposed \textit{LoGenText} \cite{ding2022logentext}, a NMT (Neural Machine Translation) approach for improving the quality of log messages: By taking the code preceding a given log statement, \textit{LoGenText} can translate it into a short textual description that can be used for logging. Such an approach can be considered complementary to the one presented in our paper.

 \smallskip 

\textbf{Log placement.} Other researchers targeted the suggestion of the best code location for log statements \cite{jia2018smartlog,li2018studying,li2020towards}. For example, Zhu \etal \cite{zhu2015learning} presented \textsc{LogAdvisor}, an approach to recommend where to add log statements. The evaluation of \textsc{LogAdvisor} on two Microsoft systems and two open-source projects reported an accuracy of 60\% when applied on pieces of code without log statements.
Yao \etal \cite{yao2018log4perf} tackled the same problem in the specific context of monitoring the CPU usage of web-based systems, showing that their approach helps developers when logging.

Li \etal \cite{li2020shall} proposed a deep learning framework to recommend logging locations at the code block level. They report a 80\% accuracy in suggesting logging locations using within-project training, with slightly worse results (67\%) in a cross-project setting. C\^andido \etal \cite{candido2021exploratory} investigated the effectiveness of log placement techniques in an industrial context. Their findings (\eg 79\% of accuracy) show that models trained on open source code can be effectively used in industry. \smallskip 

\textbf{Log level recommendation.} A third family of techniques focus on recommending the proper log level (\eg error, warning, info) for a given log statement \cite{yuan2012characterizing,oliner2012advances}. Mizouchi \etal \cite{mizouchi2019padla} proposed \textsc{PADLA} as an extension for Apache Log4j framework to automatically change the log level for better record of runtime information in case of anomalies. 
The \textsc{DeepLV} approach proposed by Li \etal \cite{li2021deeplv} uses instead a deep learning model to recommend the level of existing log statements in methods. \textsc{DeepLV} aggregates syntactic and semantic information of the source code and showed its superiority with respect to the state-of-the-art. 

Lastly, in our previous work \cite{mastropaolo2022using} we introduced \textsc{LANCE}, a tool to inject complete log statements by automatically selecting a proper log level, log message and log location.

\subsection{Combining DL and IR to Automate Code Related Tasks}
Although DL showed great potential in supporting various software engineering tasks~\cite{watsonSytematicLiterature2020}, recent work showed how its performance can be further boosted by combining it with IR-based techniques. Lam \etal~\cite{LamBugLocalization2017} proposed to use IR alongside DL for bug localization. The IR technique assesses the textual similarity between bug reports and code files. The DL model is then used to learn relationships between terms in the two different vocabularies (\ie bug reports \emph{vs} source code) and compute the final similarity score. The reported results show that DL and IR well-complement each other, with their combination outperforming the individual techniques used in isolation. Similarly, Choetkiertikul \etal~\cite{choetkiertikul2018predicting} proposed to combine IR and DL for identifying software components relevant for a given open issue.

Yu \etal~\cite{yu2022automated} combined DL with IR for the task of automated assertion generation. The idea is to use IR to retrieve the most similar test method to the target one for which an assert statement must be generated. If the similarity between the retrieved method and the target one is higher than a threshold, the assert of the retrieved method is reused. Otherwise, a DL-based approach is used to generate the assert.

In this work, we combine IR and DL to improve the performance of log statement generation, especially for what concerns the definition of a meaningful log message.
\section{Conclusions and Future Work} \label{sec:conclusions}

We started by discussing the limitations of LANCE \cite{mastropaolo2022using}, the approach we presented at ICSE'22 for the generation of complete log statements. LANCE always assumes that a \emph{single} log statement \emph{must} be injected in a method provided as input. This is a strong assumption considering that a method may not need logging or may need more than one log statement. Thus, we presented \approach, an extension of LANCE able to partially address these two limitations, making a further step ahed in the automation of logging activities. Also, we experimented in \approach a combination of DL and IR with the goal of improving the generation of meaningful log messages achieving, however, only limited improvements over LANCE.
\rev{
	In light of the results we have obtained, \approach can ensure up to 27.27\% correct predictions, when asked to inject single log statement in \java methods. On the other hand, when the model is requested to inject multiple logging statements, we observed that they were correctly added in 17\% of the methods. In addition, \approach is capable of differentiating between methods that necessitate additional log statements and those that do not, achieving an accuracy  surpassing 0.95.
}

We are working on the implementation of \approach as a tool to be deployed to developers. This is the next step needed to perform \emph{in vivo} studies, thus better understanding the main weaknesses of current DL-based log generation. 
\section{Acknowledgements}
This project has received funding from the European Research Council (ERC) under the European Union's Horizon 2020 research and innovation programme (grant agreement No. 851720).

\bibliographystyle{elsarticle-harv} 
\bibliography{main}

\end{document}